\documentclass[conference]{IEEEtran}
\IEEEoverridecommandlockouts

\usepackage{comment}
\usepackage{cite}
\usepackage{orcidlink}
\usepackage{dblfloatfix}
\usepackage{amsmath,amssymb,amsfonts}
\usepackage{algorithm}
\usepackage{algorithmic}
\usepackage{graphicx}
\usepackage{textcomp}
\usepackage{xcolor}
\usepackage{booktabs}
\usepackage{subfig}
\usepackage{multirow}
\usepackage{pgfplots}
\usepackage{plantuml}
\usepackage[nohyperlinks]{acronym}
\usepackage{hyperref}

\pgfplotsset{width=10cm,compat=1.9}
\usepgfplotslibrary{groupplots}

\def\BibTeX{{\rm B\kern-.05em{\sc i\kern-.025em b}\kern-.08em
		T\kern-.1667em\lower.7ex\hbox{E}\kern-.125emX}}

\begin{document}
	
	\title{Enabling Mixed-Criticality Applications for the Versal AI-Engines}
	
	\author{
		\IEEEauthorblockN{
			Vincent Sprave
			\orcidlink{0009-0005-8436-7065}
			\IEEEauthorrefmark{1},
			Martin Wilhelm
			\orcidlink{0000-0002-2713-2842}
			\IEEEauthorrefmark{1},
			Daniele Passaretti
			\orcidlink{0000-0001-7154-8354}
			\IEEEauthorrefmark{1},
			Alberto Garcia-Ortiz
			\orcidlink{0000-0002-6461-3864}
			\IEEEauthorrefmark{2},
			Thilo Pionteck
			\orcidlink{0000-0001-6518-1226}
			\IEEEauthorrefmark{1}
		}
		\IEEEauthorblockA{
			\IEEEauthorrefmark{1}
			\textit{Otto-von-Guericke University Magdeburg,
				39106 Magdeburg, Germany}
		}
		\IEEEauthorblockA{
			\IEEEauthorrefmark{2}
			\textit{University of Bremen,
				28359 Bremen, Germany}
		}
	}
	
	\maketitle
	\begingroup
	\renewcommand{\thefootnote}{}
	\footnotetext{%
		\textit{This paper has been accepted at the
			2026 IEEE Nordic Circuits and Systems Conference (NorCAS 2026).}
		
		\par\smallskip
		\noindent
		\copyright\ 2026 IEEE. Personal use of this material is permitted.
		Permission from IEEE must be obtained for all other uses,
		in any current or future media, including reprinting/republishing
		this material for advertising or promotional purposes,
		creating new collective works, for resale or redistribution
		to servers or lists, or reuse of any copyrighted component
		of this work in other works.
	}
	\endgroup
	\begin{abstract}
		
		Mixed-criticality systems (MCSs) are used in autonomous 
		driving, aviation and medical systems. They require 
		hardware platforms, which are reconfigurable at runtime 
		to fully utilize performance while maintaining critical 
		deadlines in emergency scenarios. However, 
		Systems-on-Chip (SoC) components, like the Versal 
		AI-Engine (AIE), only support static task mapping, 
		limiting its utilization. In this work, we propose a 
		dynamic task dispatching infrastructure that enables 
		task switching on the AIE at runtime. Based on this 
		infrastructure, we present an MCS design that 
		dynamically assigns tasks of different criticality to 
		AIE tiles, depending on the criticality mode of the 
		system. Our approach overcomes the limitations of static 
		dataflow graph mappings and, for the first time, 
		exploits the parallel processing capabilities of the AIE 
		for MCSs. We provide a timing analysis of the overhead 
		introduced by the task dispatching infrastructure, 
		which demonstrates that it is negligible. This makes 
		it suitable for MCSs, as upper bounds can be identified. 
		Finally, we evaluate the proposed infrastructure using 
		an autonomous driving workload with tasks that have 
		variable execution times and different criticality 
		levels. In this case study, we maximized AIE utilization, 
		reducing idle time by 65.5\%, while measuring an 
		execution time overhead of less than 0.8\%, and doubling 
		the throughput of low-criticality tasks.
		
	\end{abstract}
	
	\begin{IEEEkeywords}
		Heterogeneous architecture,
		Worst-Case Execution Time,
		Mixed Criticality Architecture,
		AI Engine
	\end{IEEEkeywords}
	
	% Paper sections
	\section{Introduction}

In recent years, the increasing demand for high computing performance and energy-efficient hardware in safety-critical applications, such as autonomous cars or medical devices, has driven the trend of integrating distributed computing platforms into a single heterogeneous architecture\cite{11348307,8977848}. However, this integration raises major scheduling, memory isolation and security challenges, since tasks, which have different safety, timing and resource requirements, must share the same infrastructure without interfering with each other~\cite{8101522,10.1007/978-3-031-21867-5_10}.   
Mixed-criticality systems (MCSs) address these challenges by assigning different criticality levels to tasks, thereby enabling the use of specialized scheduling and analysis methods that ensure deadlines are met~\cite{6121424}. These methods rely on worst-case execution time (WCET) estimates to provide timing guarantees, which can lead to processor underutilization and penalize low-criticality tasks. For instance, if a critical task exceeds its WCET estimate, the system may switch to a higher-criticality mode, where low-criticality tasks are suspended to guarantee that high-criticality tasks meet their deadlines~\cite{vestal_preemptive_2007}.

A heterogeneous architecture designed for computationally intensive real-time systems is AMD's Versal platform, which comprises Processing System (PS), Programmable Logic (PL),  Network-on-Chip (NoC), and AI-Engines (AIE)~\cite{10705506,11173669}. The latter consists of a processing tile matrix, but despite its high computational power, remains unexplored for mixed-criticality applications. This is because, firstly, the array's processing cores lack the ability to interrupt tasks, secondly, applications are statically mapped as dataflow graphs onto tiles. These limitations prevent task preemption and hardware reuse between high- and low-criticality tasks, limiting its integration with standard hypervisor-driven architectures, in which hardware accelerators act as co-processors for task offloading~\cite{gracioli2019designing,cinque2022virtualizing,martins2023shedding,ottaviano2024omnivisor}. The limitations of static mapping may be overcome through dynamic partial reconfiguration (DPR). However, AIE DPR has only been supported since Vivado 2024.2, and its applicability to MCSs remains unexplored. Furthermore, DPR introduces significant challenges for real-time systems, as reconfiguration may exceed $35$\,ms and can only be applied on the granularity of entire columns~\cite{AIEngineArrayReconfig}.
%due to limitations in the configuration data transfer rate~\cite{AIEngineArrayReconfig}. 
%Furthermore, DPR introduces significant challenges for communication during the reconfiguration for which there are no timing guarantee and it may exceed $35$\,ms due to the data transfer rate of configuration datawhen loading~\cite{AIEngineArrayReconfig}. 
Since reconfiguration is controlled by the PS and relies on shared resources, providing timing guarantees becomes challenging.
To overcome these limitations and to explore the processing power of the AIE in MCS, we propose a dynamic task dispatching infrastructure and a criticality-aware resource allocation strategy for the AIE. System management and task execution are done exclusively on the AIE, avoiding timing interferences with other system components. Our task dispatching strategy provides criticality-aware resource allocation, ensuring that higher-criticality tasks are prioritized. We demonstrate the feasibility of our approach by quantifying the overhead of the system  and showing that it is negligible relative to task execution times. A case study based on an autonomous driving workload confirms these results. Compared with an MCS using dedicated hardware accelerators, our approach reduces hardware idle time and allows more low-criticality tasks to be executed on the same resources.

The rest of the paper is structured as follows: Section~\ref{sec:back} introduces the fundamental concepts of MCS and the AIE microarchitecture, as well as related work.  Section~\ref{sec:MCSDesign} presents the proposed dynamic task dispatching infrastructure, and Section~\ref{sec:measurements} discusses the measurements for timing overheads. Finally, Section~\ref{sec:Proof} presents a case study.

	\section{Background}\label{sec:back}

%This section introduces the background of MCSs, the AIE architecture within the Versal SoC and related works.

\subsection{Mixed-criticality systems}
An MCS is an embedded computing platform in which application functions that share computation and/or communication resources have a different criticality based on system functionality~\cite{vestal_preemptive_2007}. Current MCS models describe the system as a set of periodic tasks $\tau$ executed sequentially. Each task is characterized by a period $P$, and must complete its execution before a deadline $D$, where $D \leq P$ is typically assumed. $D$ and $P$ are application-specific; for example, they may be derived from the required reaction time of a component to an external signal occurring at a given frequency. The task execution time $T$ depends on the allocated hardware resources and overall system utilization. To guarantee that tasks meet their deadlines, a WCET must be determined. If the WCET cannot be precisely established, a pessimistic estimate is used. Tasks are assigned a criticality level $p$ according to the impact of a potential failure, with higher-criticality tasks typically relying on more pessimistic WCET estimates. However, this pessimistic approach leads to underutilized or even non-schedulable systems.
To avoid this, multiple criticality modes $L$ are introduced~\cite{vestal_preemptive_2007}. Different WCET estimates are defined for each mode: lower-criticality modes use optimistic estimates, while higher-criticality modes use pessimistic ones. If a task exceeds the WCET estimate of a lower-criticality mode, the system switches to a higher-criticality mode, adopting a pessimistic WCET estimate. This ensures that deadlines of high-criticality tasks are met, potentially penalizing lower-criticality tasks. In this work, we consider two criticality modes --- low and high ---and therefore two execution time estimates. To clearly distinguish between them, we refer to the WCET of the low-criticality mode as the expected execution time (EET), denoted by $T_\mathrm{EET}$, and to the estimate of the high-criticality mode as the WCET, denoted by $T_\mathrm{WCET}$. Furthermore, we assume implicit deadlines, i.e., $P = D$.

\vspace{-0.2cm}
\subsection{Versal AI-Engines}

The Versal AIE is organized as a matrix of AIE tiles. Each tile contains a SIMD VLIW processor with 16 KB of program memory, an interconnect module, a dedicated Direct Memory Access (DMA) controller, and 32 KB of data memory split across eight banks. Tiles can communicate via two input and two output streams and can access both local and adjacent tile memories, providing up to 128 KB of direct memory access. The tile interconnect module contains a stream switch for AXI4-Stream traffic, and a module for handling memory mapped AXI4-traffic. Neighboring tiles within a row are also connected by a 384-bit cascade stream. The array itself is accessed over PL and NoC interface tiles.

%\begin{figure}[!h]
%%    \centering
%    \includegraphics[width=\linewidth]{figures/VersalSoC.pdf}
%    \caption{AIE array microarchitecture~\cite{AIEngineArrayReconfig}.}
%    \label{fig:aie_struc}
%    \vspace{-2mm}

%\end{figure}

Applications for the AIE are expressed as dataflow graphs, where nodes, called 'kernels', are mapped onto individual tiles. Node connections can be implemented using the communication resources of a tile. Multiple kernels can be mapped to the same tile, but cannot share resources between them and since the processor does not support preemption, they execute sequentially. For memory mapped traffic, buffers are defined and DMA controllers manage data movement between distant tiles. Memory locks enforce producer-consumer ordering and prevent data inconsistency across tiles. The Versal PS manages the start and stop operations of the graph, as well as the data copy operations over the NoC, yet it cannot alter the graph structure itself, so the topology must be fixed at compile time. This static mapping limits the AIE to functioning as an accelerator with a fixed set of static kernels and prevents kernels of different criticality levels from being dynamically reassigned to tiles at runtime.

\vspace{-0.1cm}
\subsection{Related works}
Exploiting heterogeneous architectures for MCS is an open problem that prior literature addressed, focusing on theoretical modeling, interference, data sharing, and security aspects~\cite{8101522, gracioli2019designing}. Virtualization-based approaches became the de-facto solution for running critical tasks on multiple CPUs or accelerators acting as co-processor for task-offload~\cite{cinque2022virtualizing, martins2023shedding, ottaviano2024omnivisor}.

Cinque et al.\ survey industrial virtualization practice for MCS, analyzing trade-offs in isolation, certification, and dependability~\cite{cinque2022virtualizing}. Martins and Pinto empirically compare static partitioning hypervisors for ARM-based MCS, revealing performance and safety trade-offs across Jailhouse, Xen, Bao, and seL4~\cite{martins2023shedding}. Ottaviano et al.\ extend hypervisor management to FPGA soft-cores and microcontroller-level CPUs in the Omnivisor, noting that conventional hypervisors treat such co-processors merely as I/O peripherals~\cite{ottaviano2024omnivisor}. For FPGA-specific MCS, Xia et al.~\cite{xia2019kerone} and Wulf and G{\"o}hringer~\cite{wulf_virtualization_2022} exploit DPR to share reconfigurable resources across criticality levels: Ker-ONE exposes FPGA accelerators as virtual peripherals under a preemptive hypervisor~\cite{xia2019kerone}, while L4ReC improves utilization over spatial partitioning by combining bitstream prefetching and reservation~\cite{wulf_virtualization_2022}. Gracioli et al.\ demonstrate on a Zynq UltraScale+ that hardware/software co-design is required to isolate criticality domains~\cite{gracioli2019designing}.

All of these approaches rely on the ability to reconfigure or remap hardware resources at runtime. Although parts of the AIE-array can be updated at runtime at column level via DPR starting with Vivado 2024.2, to the best of our knowledge, no existing solutions have applied DPR on the AIE for MCSs. %Furthermore, a DPR-based solution would introduce communication overhead with the PS, and timing uncertainties make it unviable for MCS.

	\section{MCS dynamic task dispatching infrastructure}\label{sec:MCSDesign}

To overcome static task mapping limitations, we propose a dynamic task-dispatching infrastructure in which resources are reallocated at runtime. Multiple tasks are mapped to the same AIE tile and controlled by a dedicated dispatcher. We also introduce a criticality-aware resource allocation strategy that enables MCS support on the AIE array despite its lack of preemptive capabilities. 

\subsection{Dynamic task dispatching infrastructure}\label{subsec:ddi}

The core component of the proposed infrastructure is a dedicated dispatcher unit that manages task execution across the AIE array. Rather than binding each task statically to a fixed compute tile, the dispatcher decides at runtime on which available compute tile an arriving task should be executed. This is possible because, through static mapping, a tile can execute multiple different tasks, allowing the dispatcher to select a tile from a pool for a specific task at runtime. A task source could be, for example, sensor interfaces from the PL, accelerators inside the PL, or data connections from the DDR.

\begin{figure}[!h]
    \centering
    \includegraphics[width=\linewidth]{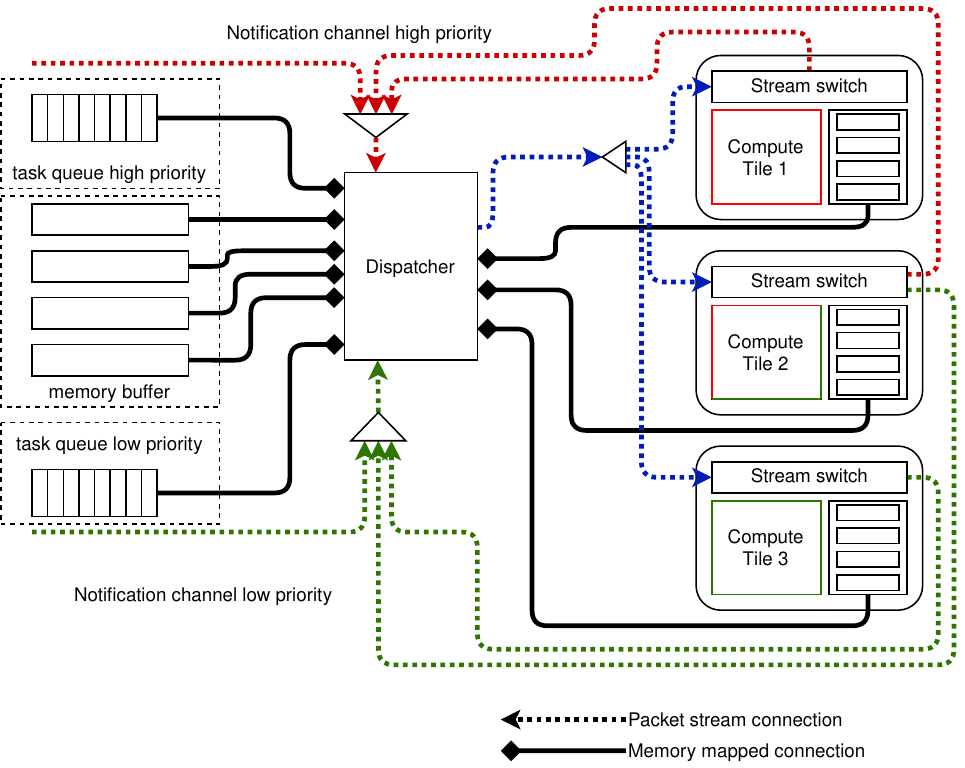}
    \caption{Infrastructure for  dynamic task dispatching.}
    \label{fig:dis}
    \vspace{-6mm}
\end{figure}
The dispatcher is implemented entirely on a single AIE tile and manages task execution, including data transfers and task activation. Task metadata, such as deadlines, task IDs, and $T_\mathrm{WCET}$, are stored in task queues, and the payload is stored in memory buffers as shown in Fig.~\ref{fig:dis}.
When the dispatcher receives a notification that a new task has arrived, it pulls the task from the queue and starts it on a free compute tile. Notifications are sent over three channels --- high-priority (red), low-priority (green), and control messages (blue) --- shown as dotted lines in Fig.~\ref{fig:dis}. They are implemented using packet streams, which allow several logical stream connections to share a single physical channel; the triangles in Fig.~\ref{fig:dis} represent the stream switches that merge or split these packet streams. Since the dispatcher has only two physical stream inputs, this merging provides a scalable solution that reduces pressure on physical resources while still allowing one input per priority level, ensuring isolation between channels of different priorities. Notification messages consist of two 32-bit packets, one header packet containing routing information for the stream switches, followed by a second packet containing the task’s unique ID. Unlike memory-mapped connections, stream interfaces support non-blocking read access, which prevents stall time in the case of channel starvation. 
This decoupling allows the dispatcher to access task metadata and workload information with minimal stall time, since AIE tile memory synchronization relies on hardware locks that grant buffer access to only one entity (processor or DMA) at a time. Because task notifications are sent only after data has been copied, the target buffer is guaranteed available upon access, avoiding stalls that would otherwise occur if the dispatcher needed a buffer still locked by the data source.

The task queue is implemented using packet streams: the first element is loaded into a memory buffer by an independent DMA controller, while the remaining elements are held in stream FIFOs whose size can be configured at design time. From the dispatcher’s perspective, the queue behaves as a buffer containing the first element, whereas from the task source perspective, it appears as a stream connection. Streams from multiple task sources are combined using stream mergers, providing a scalable design in which additional notification channels can be connected directly to the mergers.

%\begin{figure}[!h]
%    \centering
%    \includegraphics[width=\linewidth]{figures/task_queue.pdf}
%    \caption{Task queue implementation.}
%    \label{fig:taskQueue}
%    \vspace{-2mm}
%\end{figure}

In addition to task queues, buffer connections are used to copy task workload data to the task memory region, as shown by the black connections with cubic arrowheads in Fig.~\ref{fig:dis}. A task may have one or more buffer connections, typically carrying new data such as sensor data. Since task jobs are dispatched dynamically, it cannot be guaranteed that a task will have access to the same status data in the next job. For this reason, if the next job requires this status data to be consistent, it must be sent back to the dispatcher to be distributed alongside the new data. Buffer mapping plays an important role in minimizing stall time. As previously mentioned, a buffer can only be held by one entity. In hardware, however, this mechanism locks the entire memory bank, not only the memory region of the buffer. Therefore, buffers accessed by tasks that may execute in parallel must be placed in different memory banks to avoid memory contention. In contrast, buffers belonging to tasks that share the same compute tile can be placed in the same memory bank, since those tasks do not execute simultaneously.

\subsection{Criticality-aware resource allocation strategy}\label{subsec:caras}

This subsection presents the criticality-aware resource allocation strategy for MCSs, built upon the dynamic task dispatching infrastructure. Without this strategy, the dispatcher allocates resources on a best-effort basis considering only tile availability and not the deadline, which is insufficient for MCS. For this reason, the strategy bases its task allocation decisions on the current system-wide criticality mode, adapting the task pool accordingly. In a minimal configuration, as shown in Fig.~\ref{fig:dis}, the task pool consists of three compute tiles with different capabilities: tile~1 executes only high-criticality tasks, tile~2 supports both high- and low-criticality tasks, and tile~3 executes only low-criticality tasks. In low-criticality mode, high-criticality tasks are restricted to tile~1, while the remaining tiles are used for low-criticality tasks. In high-criticality mode, the execution of low-criticality tasks on shared resources (e.g., tile~2) is prohibited to ensure that sufficient resources are reserved for high-criticality tasks.

\begin{figure}[!h]
    \centering
    \vspace{-2mm}
    \includegraphics[width=0.8\linewidth]{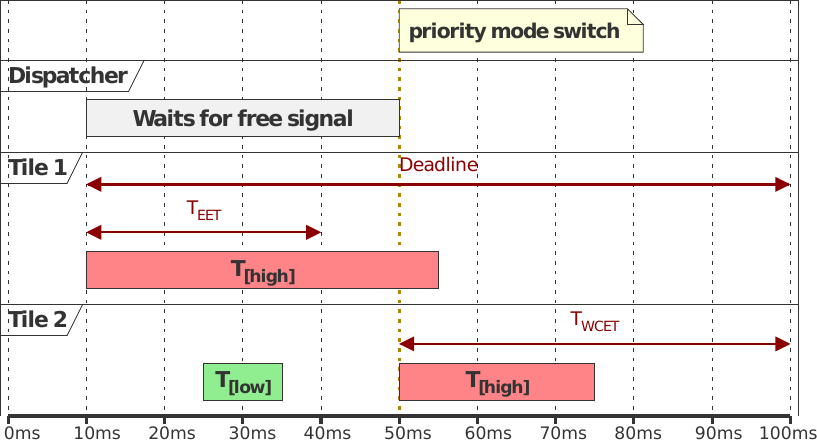}
    \caption{Timing diagram to illustrate the context switch.}
    \label{fig:scenario}
    \vspace{-3mm}
\end{figure}

This mechanism is visualized in Fig.~\ref{fig:scenario}. Initially, the system is in low-criticality mode with low-criticality tasks (green) running on tile~2 parallel with high-criticality tasks (red) on tile~1. If the high-criticality task exceeds its  $T_\mathrm{EET}$, it is not possible to guarantee the deadline of the subsequent high-criticality task  if both are assigned sequentially on tile~1. Thus, the dispatcher  switches to high-criticality mode, blocking low-criticality tasks on tile~2.

\begin{figure}[!h]
    \centering
    \vspace{-2.5mm}
    \includegraphics[width=0.9\linewidth]{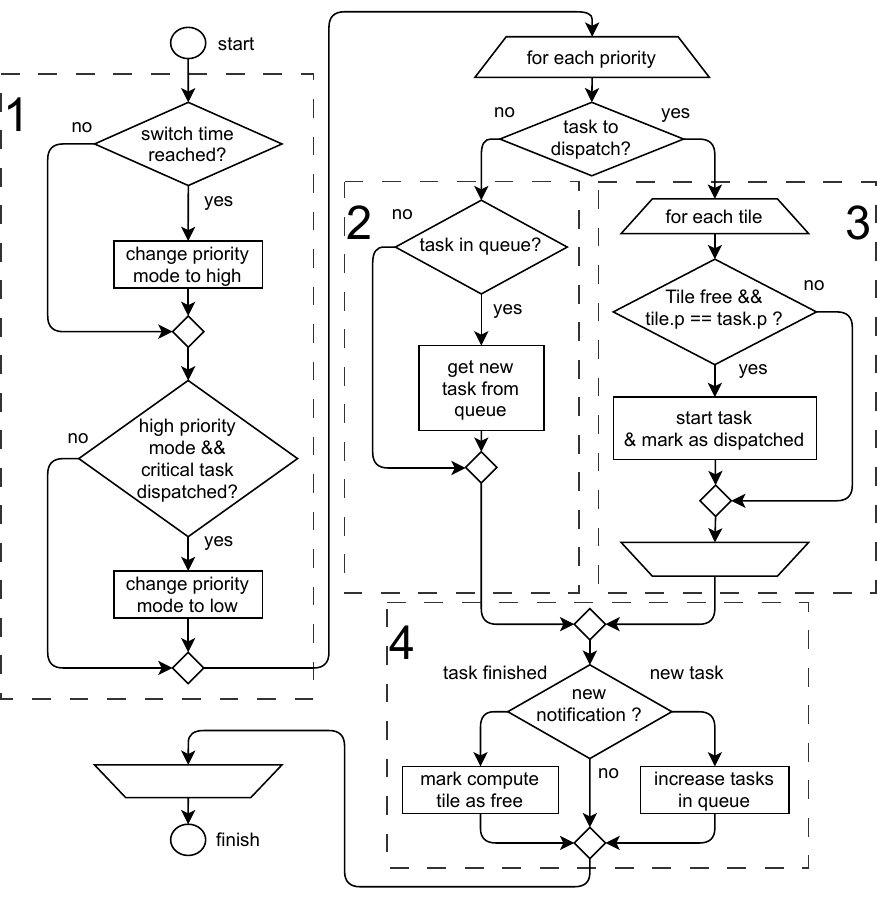}
    \caption{Criticality-aware resource allocation strategy flowchart.}
    \label{fig:flow}
    \vspace{-2mm}
\end{figure}
The flowchart of the proposed strategy is given in Fig.~\ref{fig:flow}, consisting of four steps that are executed in an infinite loop. First, the dispatcher checks whether a context switch is required, either to high-criticality mode or back to low-criticality mode. This decision is based on a timing calculation — explained in detail later — which determines the latest point in time at which the current high-criticality task must be started. If this deadline is reached, the dispatcher switches to high-criticality mode, switching back in the next iteration after this task has been successfully dispatched. The remaining three steps are executed for each priority level. As described earlier, the first element of the task queue is loaded into a buffer. To minimize access latency, the dispatcher maintains a local copy of this element. Thus, it processes one task per priority level at a time.
If the buffered task is invalid (e.g., it has already been dispatched), the dispatcher retrieves a new element from the queue (Part 2 in Fig.~\ref{fig:flow}). If the task is valid, the dispatcher attempts to assign it to a suitable compute tile (Part 3 in Fig.~\ref{fig:flow}). This is achieved by iterating over the available tiles to identify one that is both idle and capable of executing the task within the current task pool (i.e., $\text{tile}.p == \text{task}.p$). Finally, the algorithm checks whether a new notification is present in the corresponding priority-level notification channel.

The timing calculation for the context switch is detailed as follows. When a high-priority task $\tau_\mathrm{high}$ arrives, the dispatcher calculates the execution time margin by subtracting the WCET from the deadline.

\vspace{-0.3cm}
\begin{equation}
laxity = D - T_\mathrm{WCET}[\tau_\mathrm{high}]
    \label{eq:deadlin calculation}
\end{equation}
\vspace{-0.5cm}

By adding this margin to the arrival time $T_\mathrm{n,arr}$ of the task, we can determine $T_\mathrm{switch}$ which is the latest time at which a high-criticality task needs to be started. If this time is reached before the previous task finished its execution, the dispatcher performs a context switch, thus reallocating tiles used for low-criticality tasks to high-criticality tasks.

\vspace{-0.3cm}
\begin{equation}
T_\mathrm{n,switch} =T_\mathrm{n,arr} + laxity - O_\mathrm{switch}
    \label{eq:deadlineCalculation}
\end{equation}
\vspace{-0.5cm}

However, there is a delay between the context switch and reallocation and the moment when the task actually begins executing on the new tile. To account for this, the switch point must be moved earlier by this duration, which is referred to as the switching overhead, $O_\mathrm{switch}$. How this overhead time is determined will be explained in the following section.

The proposed infrastructure enables high-criticality deadlines to be met by combining low-latency packet-stream notifications with a laxity-based context switch mechanism that dynamically allocates compute tiles between criticality levels.

	\section{Measurements}\label{sec:measurements}

To guarantee that task deadlines are met, the switching overhead must be accurately determined. This section describes the methodology used to obtain timing and overhead measurements. Each AIE tile provides dedicated profiling and tracing logic that enables runtime analysis. This event logic includes a processor-independent 64-bit cycle counter, which can be accessed by the processor to timestamp relevant dispatcher events, such as task notifications. This data is streamed via dedicated channels to the DDR memory. Given the fixed AIE operating frequency of 1.25 GHz, execution times can be directly derived from the recorded cycle counts.

% change picture of timing later
\begin{figure}[!h]
   \centering
   \vspace{-1mm}
   \includegraphics[width=0.8\linewidth]{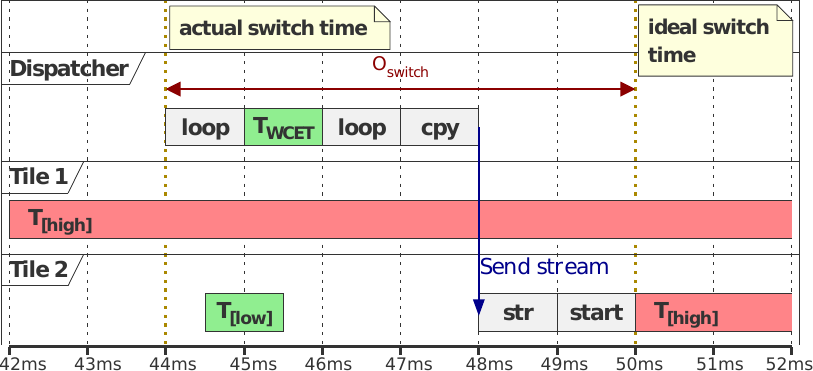}
   \caption{Timing diagram to illustrate the overhead time.}
   \label{fig:timing}
    \vspace{-2mm}
\end{figure}
% include timing paths in algorithm diagram??

Fig.~\ref{fig:scenario} illustrates the context-switch mechanism under the ideal assumption of an instantaneous switch. In practice, the switch must be scheduled earlier to account for the switching overhead $O_\mathrm{switch}$. Fig.~\ref{fig:timing} extends this scenario by showing the resulting timing behavior, with the timing intervals exaggerated for clarity. Because the dispatcher executes sequentially and does not support interrupts, the switch time cannot be precisely detected. In the worst case, this delay is one dispatcher-loop iteration, denoted by $T_\mathrm{loop}$.
After the priority mode switches, the dispatcher can schedule the high-criticality task on tile 2. However, because tasks cannot be interrupted, it must be assumed that a low-criticality task is still executing on that tile, with a worst-case execution time of $T_\mathrm{WCET}[\tau_\mathrm{low}]$. Once this task completes, an additional delay occurs before the dispatcher detects that the tile is free, which in the worst case again corresponds to one loop iteration, $T_\mathrm{loop}$. Subsequently, the task workload data must be transferred to the tile local memory, incurring a delay of $T_\mathrm{cpy}$. Finally, a notification is sent via the stream interface to start execution, introducing delay $T_\mathrm{str}$, while the tile itself requires an additional reaction time $T_\mathrm{start}$. Therefore, the switching overhead can be determined by:
\vspace{-1mm}
\begin{equation}
O_\mathrm{switch} = 2*T_\mathrm{loop} + T_\mathrm{cpy} + T_\mathrm{str} + T_\mathrm{start} + T_\mathrm{WCET}[\tau_\mathrm{low}]
    \label{eq:1}
\end{equation}
\vspace{-4mm}

 The iteration delay of the dispatcher loop, $T_\mathrm{loop}$, varies depending on whether tasks need to be dispatched. The longest time was identified by extracting the critical path in the control flow graph of the algorithm. For the tested Vitis versions (2024.2/2025.1), the cycle-accurate simulator did not properly support non-blocking streams; therefore, the individual components contributing to the path had to be simulated under worst-case assumptions and assembled. We cross-checked our analysis by hardware measurements, where runtimes of $T_\mathrm{loop}$ were collected for a scenario involving 1000 synthetic tasks, with task buffer sizes of 2~KB and 16~KB. Binary flags were used to filter the execution time measurements for iterations where the critical path was taken. The determined WCET of $T_\mathrm{loop}$ is 0.0303 ms, occurring during the loop iteration in which the task with the largest buffer is dispatched. This suggests that copying task workload data to the tile is the main contributor to execution time.
As previously mentioned, dispatcher events are recorded alongside their task IDs when the cycle counter is sampled, allowing exact start and end points of each copy operation --- and thus its duration --- to be identified precisely. 
To simplify buffer placement and memory alignment, the toolchain rounds buffer sizes up to the nearest power of two; therefore, power-of-two buffer sizes are preferred.
Following that, the copy time of buffers ranging in size from 0.5 KB to 16 KB was simulated and measured, from an input buffer on one memory bank to an output buffer on another. For small buffers, synchronization overhead dominates the total latency, whereas for larger buffers, the actual data transfer time becomes dominant. For our allocation strategy we assume a maximum buffer size of 16 KB, resulting in a fixed worst-case copy time of $0.0299$\,ms.

The control message latency $T_\mathrm{str}$ depends on the stream routing solution and can be extracted from the specification. The latency between two neighboring tiles is 8 clock cycles. As a control message consists of two packets transferred in two consecutive stream transfers, the total message latency is 9 cycles, or $7.2 \times 10^{-6}$\,ms. The reaction time $T_\mathrm{start}$ accounts for the time required by the compute kernel to read the control message (two stream reads) and compute the appropriate action before starting the task function. This time can be extracted from the measurements, with a worst-case time of $2.66 \times 10^{-4}$\,ms.
Tab.~\ref{tab:overheads} summarizes the elements of the switching overhead.

\begin{table}[!h]
\centering
\vspace{-1.5mm}
\caption{Elements of the switch overhead in ms.}
\begin{tabular}{||c c c c||} 
 \hline
 $T_\mathrm{loop}$ & $T_\mathrm{cpy}$ & $T_\mathrm{str}$ & $T_\mathrm{start}$ \\ [0.5ex] 
 \hline\hline
 $0.0303$ & $0.0299$ & $7.2 \times 10^{-6}$ & $2.66 \times 10^{-4}$  \\ 
 \hline
\end{tabular}
\label{tab:overheads}
\vspace{-2mm}
\end{table}

Using the previously described equation, the switch overhead introduced by the dispatcher (excluding $T_\mathrm{WCET}[\tau_\mathrm{low}]$) can be calculated as $0.0908$\,ms, which is rounded to $0.1$\,ms. In the following case study, the application-specific overall switch overhead (including $T_{WCET}[\tau_\mathrm{low}]$) is set to $0.2$\,ms, with $T_\mathrm{WCET}[\tau_\mathrm{low}]$ estimated to be $0.1$\,ms.

	\section{Case study: autonomous driving workloads}\label{sec:Proof}
\vspace{-0.5mm}
To demonstrate the applicability of our approach, we conducted a case study using two autonomous driving workloads for vehicle navigation: a Particle Filter (PF) task based on a LiDAR pipeline~\cite{levinson2010robust} and an FFT task for radar processing~\cite{patole2017automotive}.
Modern PFs adapt the number of particles at runtime to balance the trade-off between estimation accuracy and computational cost~\cite{fox2003kld}, which results in variable execution times. The PF was selected as the high-criticality task and the FFT task as the low-criticality representative. The task periods are based on typical sensor data rates in autonomous driving. We assume a LiDAR-based pre-processing pipeline that outputs observation points once per sensor rotation. Given typical LiDAR rotation rates of 5–20\,Hz~\cite{velodyne_hdl64e}, the PF pipeline is assumed to operate at 22\,Hz (every 45\,ms), consistent with real-time localization systems~\cite{levinson2010robust}. For the FFT task, we modeled an automotive radar~\cite{patole2017automotive} whose ADC samples the intermediate frequency beat signal at 3.4\,MHz, derived from a chirp slope of 4\,MHz/$\mu$s targeting a 150\,m range~\cite{ti_awr1843_tiduev0}. Each FFT job processes 512 complex float samples, yielding a job period of 0.15\,ms.
In our scenario, two independent PF tasks must be executed, while the remaining resources can be allocated to multiple FFT tasks. We also constrain the computation to three AIE tiles. 
%In case of multiple sensors, additional data or computation the proposed solution can be scaled using additional available AIE tiles, by instantiating additional task and dispatchers.
% In order to include these algorithms in the architecture they first have to be implemented on the AIE as individual kernels, which is shown in figure ~\ref{figs:kernel}

The PF takes two input and one output buffers. The first input buffer holds the sensor data, comprising lateral and angular velocity along with up to 12 observation points (i.e., LiDAR measurements), resulting in a buffer size of 96 bytes. The second input buffer stores the particles from the previous iteration, while the output buffer contains the updated particles, which are fed back as input for the next iteration. The number of particles per iteration is adapted based on an externally computed quality metric, with the filter dynamically selecting between 600 and 1024 particles. The particle count was capped at 1024 --- rounded up from 1000 to comply with AIE-API recommendations for power-of-two buffer sizes --- resulting in input and output buffer sizes of 16 KB. Each FFT job consumes and produces 512 complex float samples, yielding buffer sizes of 2 KB. 

To determine $T_\mathrm{WCET}$ and $T_\mathrm{EET}$, the tiles were placed in a test environment that generated input data at defined frequencies to emulate the periodic arrival of tasks such as sensor updates. The execution times could then be extracted using the cycle-accurate AIE simulator. Additionally, the results were double-checked in the running hardware setup using the measurement approach described in Section~\ref{sec:measurements}. The cycle counter was extracted at the start and end of each job, and both timestamps were transmitted to the DDR memory via a dedicated stream. This method is sufficient, as the tile processor is deterministic and the execution time on the tile itself varies little over multiple job iterations. A total of 1000 jobs were measured for each task. 
For the FFT task, a maximum execution time of 0.096 ms was observed. As this task executes the same number of samples each time, it is assumed that $T_\mathrm{WCET} = T_\mathrm{EET}$. When the PF algorithm chooses a lower particle count, a single job executes for $T_\mathrm{EET}$. A maximum execution time of 14.68 ms was observed in this case. Conversely, $T_\mathrm{WCET}$ is observed when the algorithm utilizes the maximum amount of particles, resulting in a maximum execution time of 24.79 ms.  From these measurements, the $T_\mathrm{WCET}$ and $T_\mathrm{EET}$ of both task were estimated and are summarized in Tab.~\ref{tab:2}.

% \begin{figure}[htbp]
% \vspace{-1mm}
% \centering
% \begin{tikzpicture}

% \pgfplotstableread[col sep=comma]{data/runtimePF.csv}\dataPF
% \pgfplotstableread[col sep=comma]{data/runtimeFFT.csv}\dataFFT

% \begin{groupplot}[
%     group style={group size=1 by 2, vertical sep=1.2cm},
%     ybar,
%     ymin=0,
%     height=3.5cm,      
%     width=9cm,
% ]

% % ---- First plot ----
% \nextgroupplot[        
%     xlabel={Runtime of a single PF-task [ms]},xtick={13,14,15,16,17,18,19,20,21,22,23,24,25},
%     bar width=2pt
% ]

% \addplot table[x=right, y=count] {\dataPF};

% % ---- Second plot ----
% \nextgroupplot[        
%     xlabel={Runtime of a single FFT-task [ms]},xmin=0,
%     xmax=0.2,
%     xtick={0,0.05,0.1,0.15,0.2},
%     xticklabel style={/pgf/number format/fixed},
%     bar width=2pt
% ]

% \addplot table[x=right, y=count] {\dataFFT};

% \end{groupplot}
% \node[rotate=90]
%     at ($(group c1r1.west)!0.5!(group c1r2.west)$)
%     [above=0.75cm]
%     {Number of measurements};
% \end{tikzpicture}
% \caption{Particle filter and FFT execution time measurements.}
% \label{fig:runtime}
% \vspace{-1mm}

% \end{figure}
%

\begin{table}[!h]
\caption{Summary of task parameters.}
\vspace{-0.5mm}
\centering
\begin{tabular}{||c c c c c||} 
 \hline
 Task & Priority & $T_\mathrm{WCET}$ & $T_\mathrm{EET}$ & $D = P$  \\ [0.5ex] 
 \hline\hline
 Particle Filter & High & 25 ms & 15 ms & 45 ms  \\ 
 \hline
 FFT & Low & 0.1 ms & 0.1 ms & 0.15 ms  \\
 \hline
\end{tabular} 
\label{tab:2}
\vspace{-2mm}
\end{table}

In a static mapping approach, it must be assumed that a high-criticality task will exceed its $T_\mathrm{EET}$ with 100\% probability, since varying execution times cannot be dynamically compensated for. Consequently, two high-criticality tasks cannot be assigned to the same tile, which results in underutilization during low-criticality operation. In practice, however, the probability of exceeding $T_\mathrm{EET}$ is lower. This increases the benefit of the proposed infrastructure since resources are allocated dynamically based on the current criticality mode. Consequently, additional execution time is available for lower-criticality tasks. For the following analysis, a pessimistic probability of 20\% is assumed.

\begin{figure}[!h]
   \centering
   \vspace{-2mm}
   \includegraphics[width=0.7\linewidth]{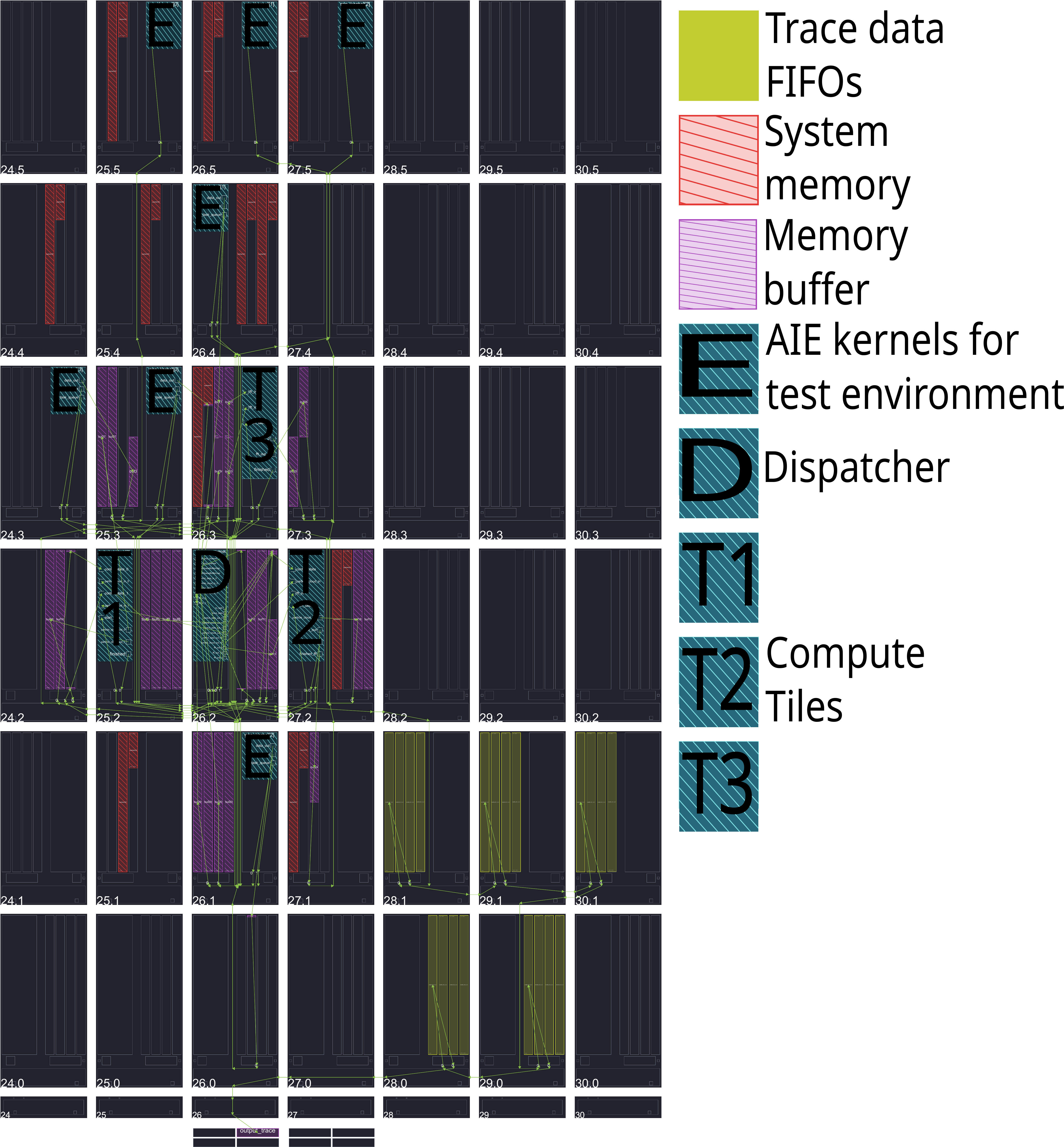}
   \caption{Mapping solution of the demonstrator.}
   \label{fig:caseStudy}
   \vspace{-2mm}
\end{figure}

The demonstrator was implemented on the XCVC1902 Versal SoC and the mapping is shown in Fig.~\ref{fig:caseStudy}. For the criticality-aware resource allocation, the tile usage is defined as follows. In low-criticality mode, high-criticality tasks are dispatched exclusively to T1, while the remaining tiles are reserved for low-criticality tasks. In high-criticality mode, high-criticality tasks may be dispatched to both T1 and T2, while tile T3 is reserved for low-criticality tasks.

\begin{figure}[!h]
    \centering
    \vspace{-3mm}
    \begin{tikzpicture}

        % Read both CSV files
        \pgfplotstableread[col sep=comma]{data/tileUsageStatic.csv}\dataone
        \pgfplotstableread[col sep=comma]{data/tileUsageArch.csv}\datatwo

        \begin{groupplot}[
            group style={
                group size=2 by 1,
                horizontal sep=0.8cm
            },
            ybar stacked,
            width=7cm,
            height=6cm,
            enlarge x limits=0.2,
            symbolic x coords={T1,T2,T3},
            xtick=data,
            xlabel={},
            ytick={0,4.5,9},
            cycle list={red,purple,blue,orange,gray},
            every axis plot/.append style={fill},
            legend to name=executionlegend,
            legend style={draw=none, legend columns=1}
        ]

        % --- First stacked bar chart ---
        \nextgroupplot[
        title={Static Mapping},
        ylabel={Exec Time [s]},
        title style={yshift=-0.2cm},
        scale=0.45
        ]
        \addplot table[x=Tile,y=High1] {\dataone};
        \addplot table[x=Tile,y=High2] {\dataone};
        \addplot table[x=Tile,y=Low1] {\dataone};
        \addplot table[x=Tile,y=Low2] {\dataone};
        \addplot table[x=Tile,y=Free] {\dataone};

        % --- Second stacked bar chart ---
        \nextgroupplot[
        title={Our Architecture},
        title style={yshift=-0.2cm},
        scale=0.45
        ]
        \addplot table[x=Tile,y=High1] {\datatwo};
        \addplot table[x=Tile,y=High2] {\datatwo};
        \addplot table[x=Tile,y=Low1] {\datatwo};
        \addplot table[x=Tile,y=Low2] {\datatwo};
        \addplot table[x=Tile,y=Free] {\datatwo};
        \legend{High1, High2, Low1, Low2, Idle}

        \end{groupplot}

        \node[
            anchor=west,
            xshift=0.5cm
        ] at (group c2r1.east) {\ref{executionlegend}};

    \end{tikzpicture}
    \vspace{-6mm}
    \caption{Execution time distribution comparison.}
    \label{fig:executionTimes}
\vspace{-6mm}
\end{figure}

 Fig.~\ref{fig:executionTimes} shows the execution time distribution of the proposed architecture compared to a static mapping approach over a period of nine seconds. The dynamic dispatching infrastructure makes more efficient use of available computing resources. If both PFs stay in their $T_\mathrm{EET}$, T2 can be used for additional low-criticality tasks, doubling the number of low-criticality tasks executed compared to static mapping. This confirms the advantages of the proposed infrastructure and the strategy for MCS, reducing the overall idle time by 65.5\% compared to static mapping. The infrastructure introduces only an overhead of 0.2 ms, which corresponds to less than 0.8\% of the $T_\mathrm{WCET}$ of the high-criticality task.

This case study demonstrates the applicability of the proposed infrastructure in the context of MCS for a four-task workload mapped onto a minimal system configuration consisting of three compute tiles. However, this is intended as a proof of concept, and the architecture can be scaled up for larger systems. One approach would be to instantiate the architecture multiple times for different regions of the AIE matrix. For a single instance, additional compute tiles can be integrated by extending the notification channels through the stream split and merger functions. In practice, scalability is constrained by memory contention arising from an increasing number of buffer connections, this can be mitigated by introducing dedicated tiles for memory distribution. The number of tasks mapped to a compute tile can also be increased, limited mainly by tile data and program memory capacities. %Program memory usage may be optimized by replacing complex computations with look-up tables.

	\section{Conclusion}\label{ref:concl}

We presented a dynamic task dispatching infrastructure and a criticality-aware resource allocation strategy that, for the first time, enable MCSs on the Versal AIE array. By mapping multiple tasks to shared compute tiles and leveraging packet-stream-based notification channels with a laxity-driven context switch mechanism, our approach overcomes the fundamental limitations of the AIE's static dataflow graph model without relying on task preemption or DPR. A case study on an autonomous driving workload comprising PF and FFT tasks validates the approach, showing a 65.5\% reduction in tile idle time and doubling the low-criticality task throughput compared to static mapping, while all high-criticality deadlines are met. Timing analysis confirms that the dispatcher overhead remains below 0.8\% of the WCET of the high-criticality task, demonstrating its suitability for real-time operation.
	
	\section*{Acknowledgements}
	
	This project is funded by the German Research Foundation
	(Deutsche Forschungsgemeinschaft, DFG) -- 524884424
	
	% IEEE bibliography with numerical citations
	\bibliographystyle{IEEEtran}
	\bibliography{references}
	
\end{document}